\documentclass{emulateapj}

\usepackage{graphicx}
\usepackage{epstopdf}

\newcommand{\flux}{erg~cm$^{-2}$s$^{-1}$}
\newcommand{\lx}{erg~s$^{-1}$}

\newcommand{\chandra}{{\it Chandra}}
\newcommand{\xmm}{{\it XMM-Newton}}
\newcommand{\rsun}{$R_{\odot}$}
\newcommand{\msun}{$M_{\odot}$}
\newcommand{\degree}{$^{\circ}$}

\slugcomment{Submitted to MNRAS. Manuscript Printed Oct 11, 2014}

\shorttitle{IC 10 X-1}
\shortauthors{Laycock et al.}

\begin{document}

\title{Chandra and XMM Monitoring of the Black Hole X-ray Binary IC 10 X-1}


\author{Silas G. T.  Laycock, Rigel C. Cappallo, and Matthew J. Moro}
\affil{Department of Physics and Applied Physics, Olney Science Center, University of Massachusetts Lowell, MA, 01854, USA}

\label{firstpage}



\begin{abstract} 

The massive black hole + Wolf-Rayet binary IC10 X-1 was observed in a series of 10 \chandra ~and 2 \xmm ~observations spanning 2003-2012, showing consistent variability around 7$\times10^{37}$ \lx, with a spectral hardening event in 2009. We phase-connected the entire light-curve by folding the photon arrival times on a series of trial periods spanning the known orbital period and its uncertainty, refining the X-ray period to P = 1.45175(1)d. The duration of minimum-flux in the X-ray eclipse is $\sim$5 hr which together with the optical radial velocity curve for the companion yields a radius for the eclipsing body of 8-10\rsun ~for the allowed range of masses. The orbital separation $(a_1+a_2)$ = 18.5-22\rsun ~then provides a limiting inclination $i>63$\degree ~for total eclipses to occur. The eclipses are asymmetric (egress duration $\sim0.9$hr) and show energy dependence, suggestive of an accretion-disk hotspot and corona. The eclipse is much ($\sim$5X) wider than the 1.5-2\rsun WR star, pointing to absorption/scattering in the dense wind of the WR star. The same is true of the close analog NGC 300 X-1. RV measurements of the He II [{\small $\lambda\lambda$}4686] line from the literature show a phase-shift with respect to the X-ray ephemeris such that the velocity does not pass through zero at mid-eclipse. The X-ray eclipse leads inferior conjunction of the RV curve by $\sim$90\degree, so either the BH is being eclipsed by a trailing shock/plume, or the He II line does not directly trace the motion of the WR star and instead originates in a shadowed partially-ionized region of the stellar wind. 
\end {abstract}


\section{Introduction}

The extra-galactic X-ray binary IC10 X1 consists of a black hole (BH) and a wolf-rayet (WR) star orbiting their common centre of mass with a period of 34.9 hrs.  X-ray timing and optical radial velocity studies (\citealt{Prestwich2007}, \citealt{SF}) have determined a range of dynamically allowable masses for the two objects. The lowest mass scenario is $M_{*}$ = 17 \msun,  $M_{BH}$ = 23 \msun, and the most massive scenario is   $M_{*}$ = 35 \msun,  $M_{BH}$ = 32 \msun. These configurations follow from the characterization of the Wolf-Rayet star by \cite{Clark2004} and the mass function determined from the radial velocity curve of the primary  \citep{SF}. Deep eclipses are seen in the X-ray light-curve as was first noted by \cite{Prestwich2007}, who confirmed this fact with a 10 day series of Swift observations in 2006.  The system resides in the nearby dwarf starburst galaxy IC 10 at a distance of 660 kpc \citep{Wilson1996}.  

In this article we describe the X-ray properties of IC 10 X-1 over the period 2003-2014, focussing on the \chandra ~dataset, and incorporate the  \xmm ~broad-band light-curves in order to extend the duration of the time series.

IC10 X-1 belongs to the small group of known Black Hole High Mass X-ray binaries (BH-HMXBs), which includes Cyg X-1 (\citealt{Bolton1972}, \citealt{Hutchings1973}), Cyg X-3 \citep{Hanson2000},  LMC X-1 \citep{Orosz2009}, LMC X-3, \citep{Nowak2001}, M33 X-7 \citep{Orosz2007}, NGC 300 X-1 (\citealt{Carpano2007}, \citealt{Crowther2010}, \citealt{Binder2011}), and CXOU J123030.3+413853 in NGC4490 \citep{Esposito2013}.  We note the incidence of apparently eclipsing systems is surprisingly high among BH-HMXBs.

The majority of known stellar black holes and candidate systems are in low mass systems (LMXBs), probably because the lifetimes of lower-mass donor stars are longer. In contrast, the most massive stellar BHs appear to have the most massive companions; the current record-holders are M33 X-7 (a 70 \msun O-star) and IC 10 X-1. This phenomenon that has been investigated by \cite{deMink} who show that rapid rotation and tidally-induced mixing between the core and envelope alters the evolution of extremely massive binary stars in an interesting way. The companion avoids the red-giant phase and evolves directly to a massive helium star (effectively a WR star), retaining a much higher mass, and a more compact size than in the isolated case. Detailed binary evolutionary scenarios have been explored by \cite{Bogomazov2014} who note that their birth events are leading candidates for long gamma ray bursts (GRBs), and that BH-HMXBs' eventual fate as BH+BH binaries will be a significant source of gravitational wave radiation. We suggest \cite{McClintock2006} and \cite{Frolov2011} for comprehensive reviews of observations and theory of BH X-ray binaries. 

A key goal in this study was to refine the binary period to the point that its derivative ($\dot{P}$) could be tracked by future observations. The rate at which the orbital period changes ($\frac{\dot{P}}{P}$) is a crucial parameter in BH binary systems, as it contains measurable contributions from gravitational radiation, magnetic braking, and mass-loss from the primary \citep{Gonzalez2014}. Fundamental research in general relativity and particle physics requires more such measurements; for example \cite{Johannsen2009} show that $\dot{P}$ in BH binaries provides one of the only experimental constraints on the properties of extra dimensions predicted by string theory. Evolutionary simulations of BH-HMXBs by \cite{Tutukov2013} suggest that experimental determination of $\frac{\dot{P}}{P}$ in IC 10 X-1 is the most promising route to understand the relationship between the companion's wind properties and the X-ray luminosity.

Improving the orbital ephemeris is crucial for other reasons too. Firstly, for interpretation of the existing data it is necessary to know the orbital phase; for example to generate the eclipse profile, or to compare the spectrum in and out of eclipse. Secondly, no existing radial velocity studies for IC 10 X-1 and NGC 300 X-1 have explicitly demonstrated that the HeII $\lambda\lambda$4686 emission line shift passes through zero at X-ray minimum \citep{Tutukov2013}. The HeII line emission used to obtain the RV curve is probably inhomogeneous and variable, hence it is not certain that the radial-velocity curves of the optical stars in these systems accurately reflect their orbital motions.  

The X-ray spectrum of IC10 X-1 has been characterized from \xmm ~and \chandra ~observations of 2003 by \cite{Wang2005} and \cite{BB}. Both groups found a spectrum that is similar to other BH-HMXBs, exhibiting a hard  accretion-disk component embedded in a softer thermal extended emission. 

In all published light-curves of X1 the source exhibits flaring on timescales of minutes to hours, and several observations show slower variations which are likely segments of eclipse ingress/egress, but could plausibly be due to variations in the mass transfer rate.  A  $\sim$7 mHz quasi-periodic oscillation (QPO) has been discovered in the X-ray light-curve by \cite{Pasham2013}. Its frequency is similar to that seen in highly inclined X-ray binaries; in which the accretion disc obscures the central source leading to so-called Dipping behavior. Alternatively ``Heartbeat" QPOs due to radiation pressure instability of the inner edge of the accretion disk can also appear in the same frequency range. Differentiation of these models could come from additional QPO measurements as the system transitions to different accretion states. Luminosity variations occur in BH-HMXBs on all timescales, as is seen in for example Cyg X-1 \citep{Grinberg2014}, thus establishing the long-term activity of IC 10 X-1 and occurrence of state-transitions is essential to understanding the stability of its accretion geometry.

\section{Observations and Data Reduction}

IC 10 has been observed by \chandra ~10 times and twice by \xmm ~ as listed in Table~\ref{tab:dataset}. The monitoring series of 7 observations (late 2009 through 2010) were intended to fill-in the phase coverage of IC 10 X-1's orbit, extend the baseline in order to refine the orbital period, and search for transient X-ray sources \citep{Laycock2010,Laycock2014}. \chandra ~data reduction was performed in {\it Ciao}\footnote{http://cxc.harvard.edu/ciao} following the standard processing threads for point-sources. The \chandra ~ACIS-S Level-2 event files were barycentre corrected with {\it abary} to remove timing artifacts due to Earth's motion around the Sun.  For the declination of IC 10 (+59.3 \degree) the annual path-length variation is $\pm$254 light seconds. Source fluxes were computed in 3 bands (broad-band B = 0.3-8 keV, soft-band S = 0.1-1.5 keV, and hard-band H = 5.0-10 keV) from {\it wavdetect} output, and the same bands were used to generate light-curves and hardness ratios. Event files were extracted for the source using the 95\% encircled energy radius. Background events were extracted in an annulus bounded by the corresponding 3$\sigma$ and 4$\sigma$ radii.  Exposure maps generated for 1.5 keV were used to correct for sensitivity variations.  

\begin{deluxetable}{lllllll}
\tablecaption{X-ray Observations of IC10 X-1
 \label{tab:dataset}} 
\tablehead{ \colhead{MJD}  &   \colhead{Date} &  \colhead{Inst.}  &   \colhead{ObsID}         & \colhead{Offset}     &   \colhead{Exp} & \colhead{Cts}  \\
                                                    &                     &                                    &                                         &   \colhead{\scriptsize arcmin} &      \colhead{\scriptsize ksec}       &     \colhead{$\times10^{3}$}       }
\startdata
52710.7 &  12 Mar 2003   & CXO  & 03953 & 0.48 & 28.9  &  5.21     \\
52823.9  & 2 July 2003   & XMM      &  152260              &   0.02      &   42    &   21  \\ 
 54041.8 & 2 Nov 2006  & CXO & 07082 & 3.20  & 40.1 &   3.75 \\
54044.2 & 5 Nov 2006    & CXO & 08458 & 3.20 & 40.5  &  4.58     \\
55140.7 &  5 Nov 2009   & CXO & 11080 &1.90  & 14.6   &  0.54  \\
55190.2 &  25 Dec 2009  & CXO & 11081 &1.74  & 8.1  &   0.9     \\
 55238.5 & 11 Feb 2010  & CXO &11082 & 0.83 & 14.7   &  2.14    \\
 55290.6 & 4 Apr 2010    & CXO &11083 &2.25 & 14.7    &   2.33     \\
55337.8 & 21 May 2010 & CXO &11084 &  3.45 & 14.2     &   0.56   \\
55397.5 & 20 Jul 2010   & CXO &11085 & 2.27 & 14.5    &  0.34    \\
55444.6 & 5 Sep 2010   & CXO &11086 & 2.22 & 14.7   &    2.26 \\
 56157.9 & 18 Aug 2012 & XMM       &  693390          &   0.01        &  135   &  48.6   \\
\enddata
\tablecomments{Instrument is \chandra ~ACIS-S3 or \xmm ~PN. Cts is net background subtracted counts} 
\end{deluxetable}

Source and background spectra with region-specific response files (RMF, ARF) were created using the {\it Ciao} script {\it specextract} for each observation. The data and response files were analysed in XSPEC 12.7\footnote{http://heasarc.nasa.gov/xanadu/xspec}.  

Further analysis of all light-curves and events was performed in {\em R}\footnote{The R-Project for Statistical Computing. www.r-project.org}

\xmm ~light-curves were obtained from the standard pipeline products in the \xmm ~Science Archive (XSA). We used the background corrected broad-band PN light-curve at 40 s time resolution barycentre corrected for Earths motion.

\section{Light-curves: Flux and Hardness Ratio Variability}
\begin{figure}
\begin{center}
\includegraphics[trim= 0 0cm 0 1cm, clip,angle=0,width=9cm]{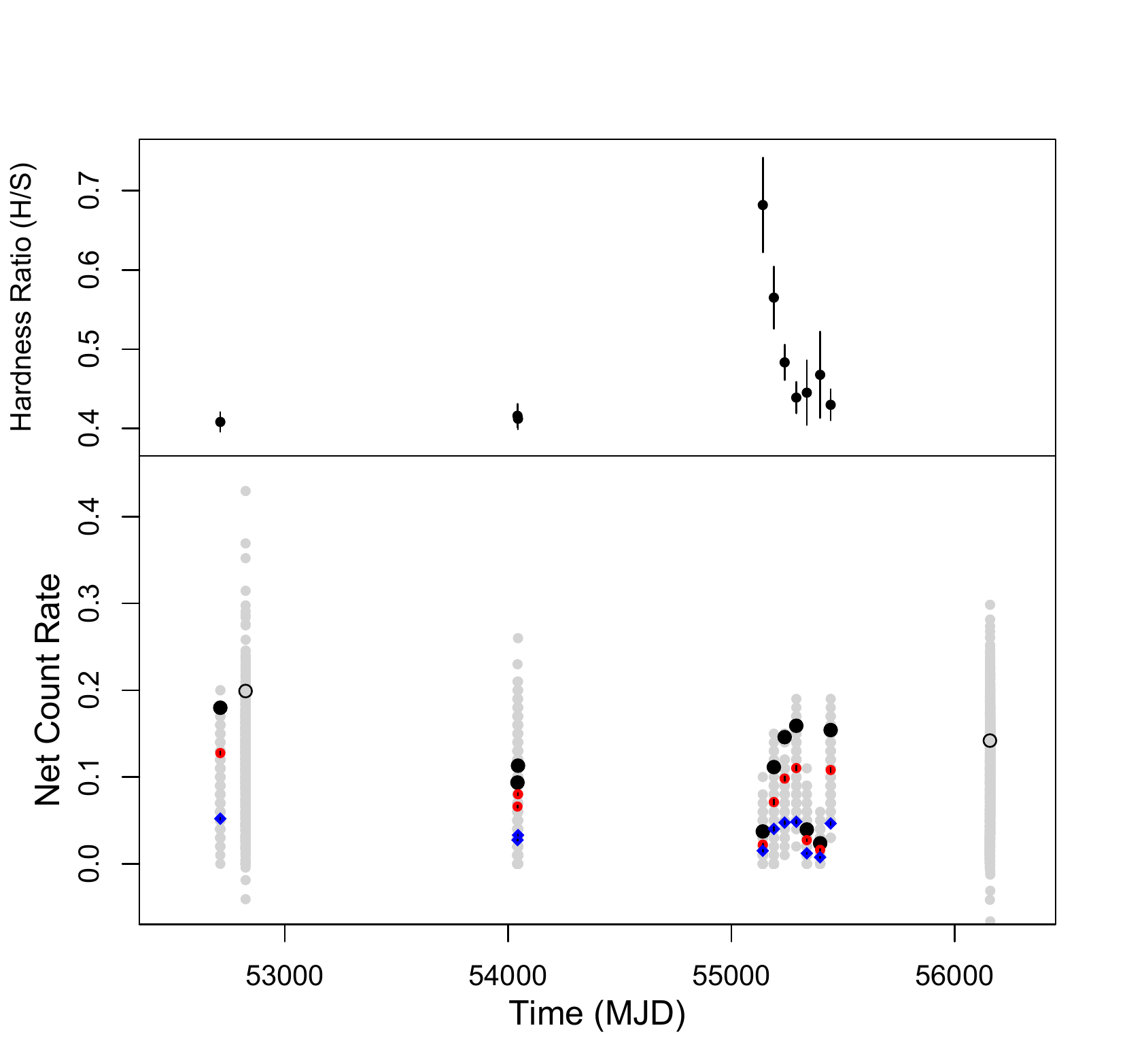}
\caption{{\bf  Long-Term X-ray Light-curve and Hardness Ratio for IC 10 X-1. Lower panel: Average \chandra ~ACIS-S count-rates for each observation are plotted in 3 energy bands: B=0.3-8 keV (black), S=0.3-1.5 keV (red), and H=2.5-8 keV (blue). The two XMM-Newton PN measurements (open circles) were translated to the ACIS-S B-band using PIMMS. During all observations the count rate varied strongly due to a combination of eclipses and flaring, this is indicated by the grey points which are broad-band values obtained at 100 sec resolution during each observation. Upper panel: Hardness Ratio (H/S) for the ACIS-S observations with statistical error bars. }}
\label{fig:longlc}
\end{center}
\end{figure}

X1 is detected in all 10 observations of IC 10, and the resulting long-term light-curve from \chandra ~and \xmm ~is plotted in Figure~\ref{fig:longlc}. The \chandra ~count-rates have been corrected for exposure time, exposure map (i.e. spatial location on the detector) and changes in detector sensitivity with time. The \xmm ~points have been converted to the same flux scale as \chandra ~using WebPIMMS. The observations vary in integration time (Table~\ref{tab:dataset}) and the intrinsic source flux varies significantly during each exposure. The solid points in the long-term light-curve are the average flux measured during each observation, the range of variability is shown by light grey dots. The broad-band count-rates in Figure~\ref{fig:longlc} show a factor of $\sim$4 variability both between and within observations. The variability contains contributions from eclipses, intrinsic source variability (flickering), and possibly changes in accretion rate and geometry.

The hardness ratios (HR=H/S) reveal that a change in spectral shape took place during the 2009-10 monitoring series. Over the course of 4 observations the HR dropped from a peak of 0.68$\pm$0.06 back to 0.44$\pm$0.02, which is consistent with its prior (2003, 2006) value of 0.41$\pm$0.015. The peak value is an increase of 67\% over the baseline value, while at the same time the amplitude of the short-term (intra-observation) flux variability appears to shrink. 

\begin{figure*}
\begin{center}
\includegraphics[trim= 0 1cm 0 0, clip,angle=0,width=18cm]{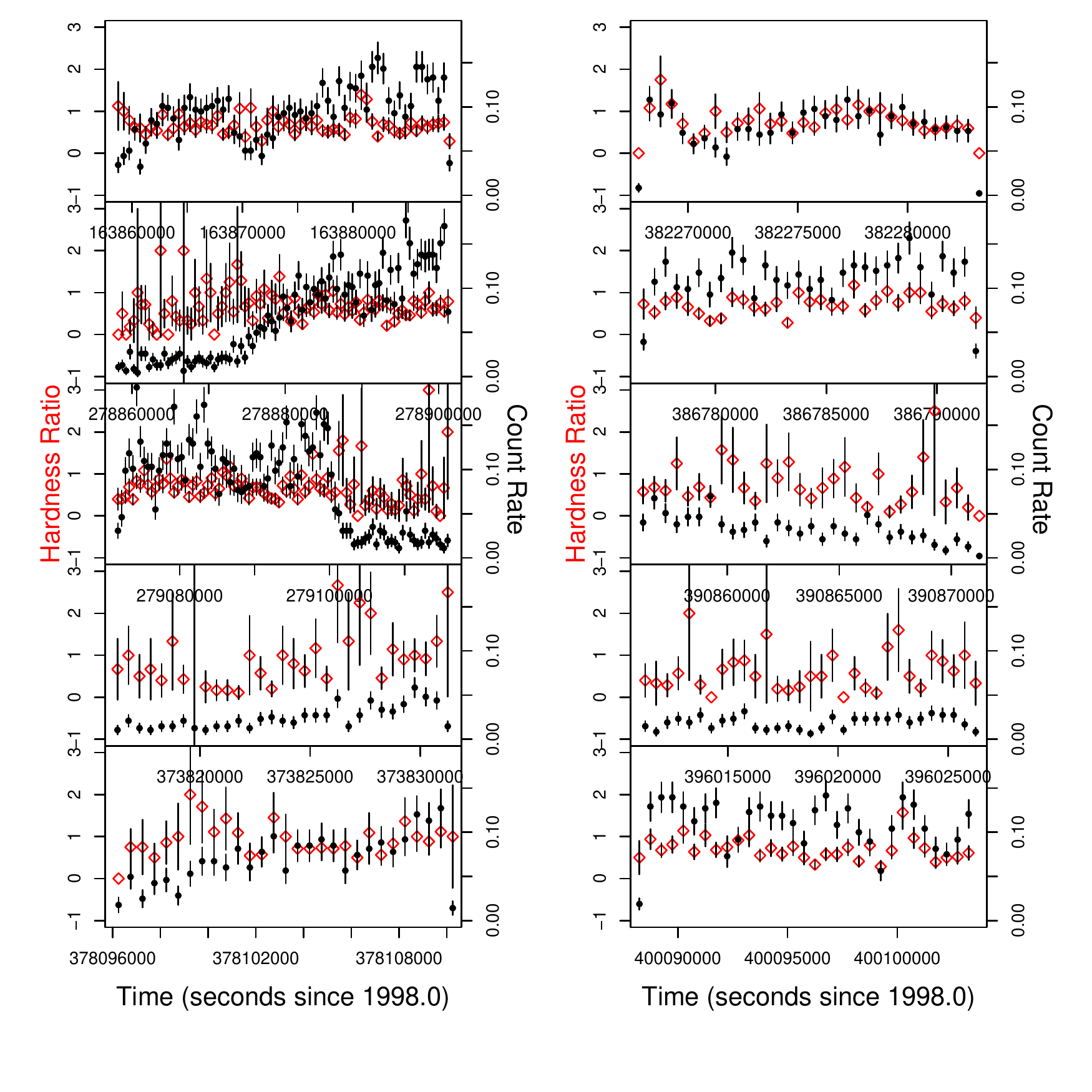}
\caption{{\bf  Light-curves and Hardness Ratios at 500 sec resolution for all 10 Chandra observations. All plots share the same Y-axes: net count-rates (corrected for exposure map) are plotted as solid dots, with scale on the right-hand y-axis, hardness ratios are red diamonds with scale on left Y-axis. The X-axes are chronological sections of the mission timeline (seconds since 1998.0 TT). The second and third panels (ObsIDs 07082 \& 08458) contain respectively the clearest examples of eclipse egress and ingress.  }}
\label{fig:rate_and_hr_all}
\end{center}
\end{figure*}

We examined variability within each \chandra ~observation by plotting broad-band count-rate (CR) and hardness ratio (HR) at 500 sec resolution, as displayed in Figure~\ref{fig:rate_and_hr_all}.  The most obvious features are the eclipse egress and ingress caught in observations 2 \& 3. Even during eclipse the broad-band flux does not drop to zero, and variability is always present. This suggests that the emission region is not fully eclipsed and that the emission is not point-like. A point source undergoing eclipses produces an instantaneous change in flux, whereas the \chandra ~CR in Figure~\ref{fig:rate_and_hr_all} (Panels 2 \& 3) takes 1-2 hours to fully transition.

Within all observations the HR fluctuates by a smaller proportion than the broad-band CR. Most of the time the HR appears generally positively correlated with CR, such as in \chandra ~observations 6, 7 \& 8. At other times large changes appear in the CR that do not register in HR (e.g. the entire latter half of observation 2), or appear only as features of much shorter duration; for example 3/4 of the way through observation 1.

\begin{figure}
\begin{center}
\includegraphics[trim= 0 4cm 0 2cm, clip,angle=0,width=9cm]{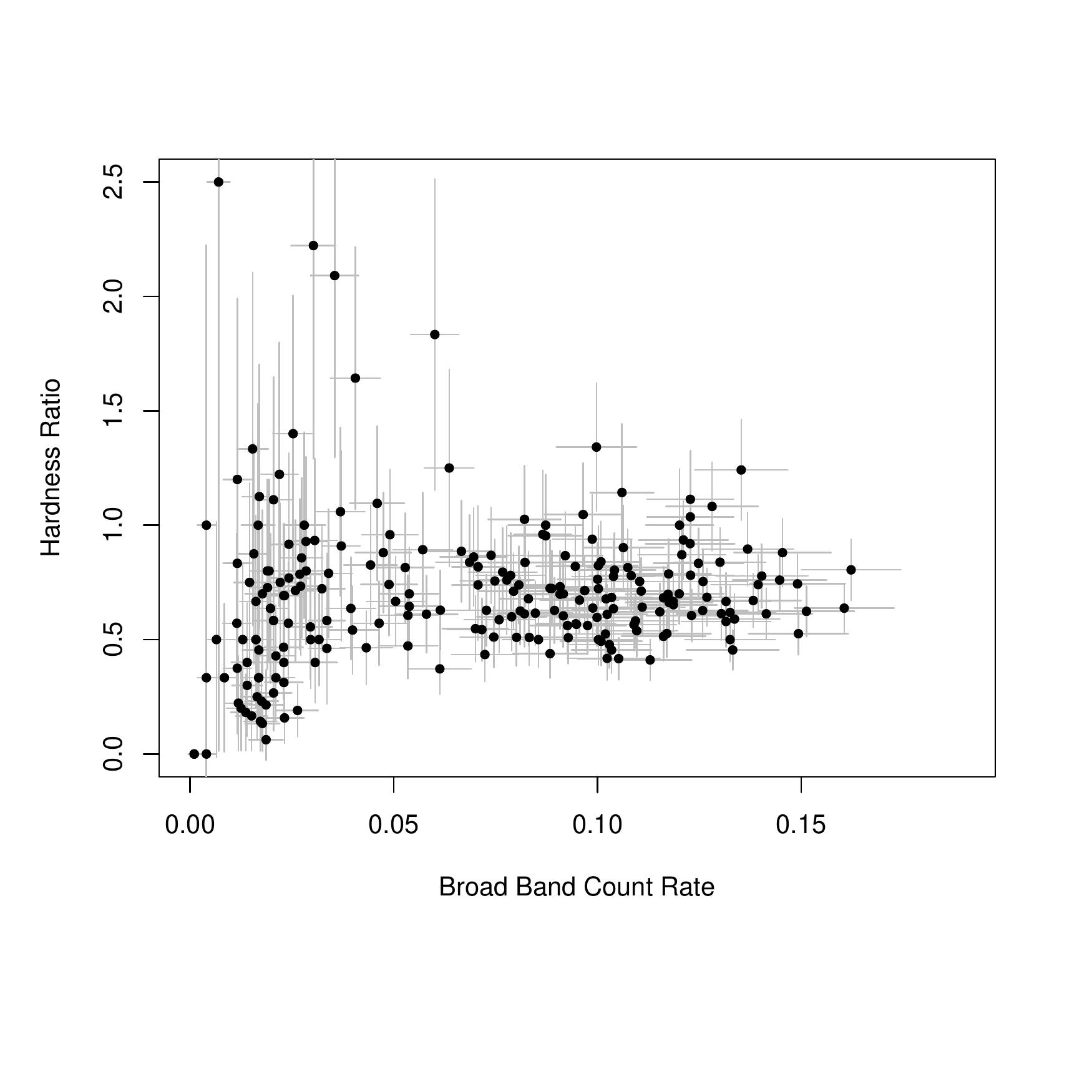}
\caption{{\bf  X-ray Count-Rate vs Hardness Ratio diagram for the combined Chandra data. The net count-rate is in the broad 0.3-8 keV band, the HR is constructed from the ratio of Hard band  / Soft band in 1000 sec bins.  There is a well defined split between high and low flux values, occurring at CR = 0.06.}}
\label{fig:chandra_CMD}
\end{center}
\end{figure}

\begin{figure}
\begin{center}
\includegraphics[trim= 0 4cm 0 2cm, clip, angle=0,width=9cm]{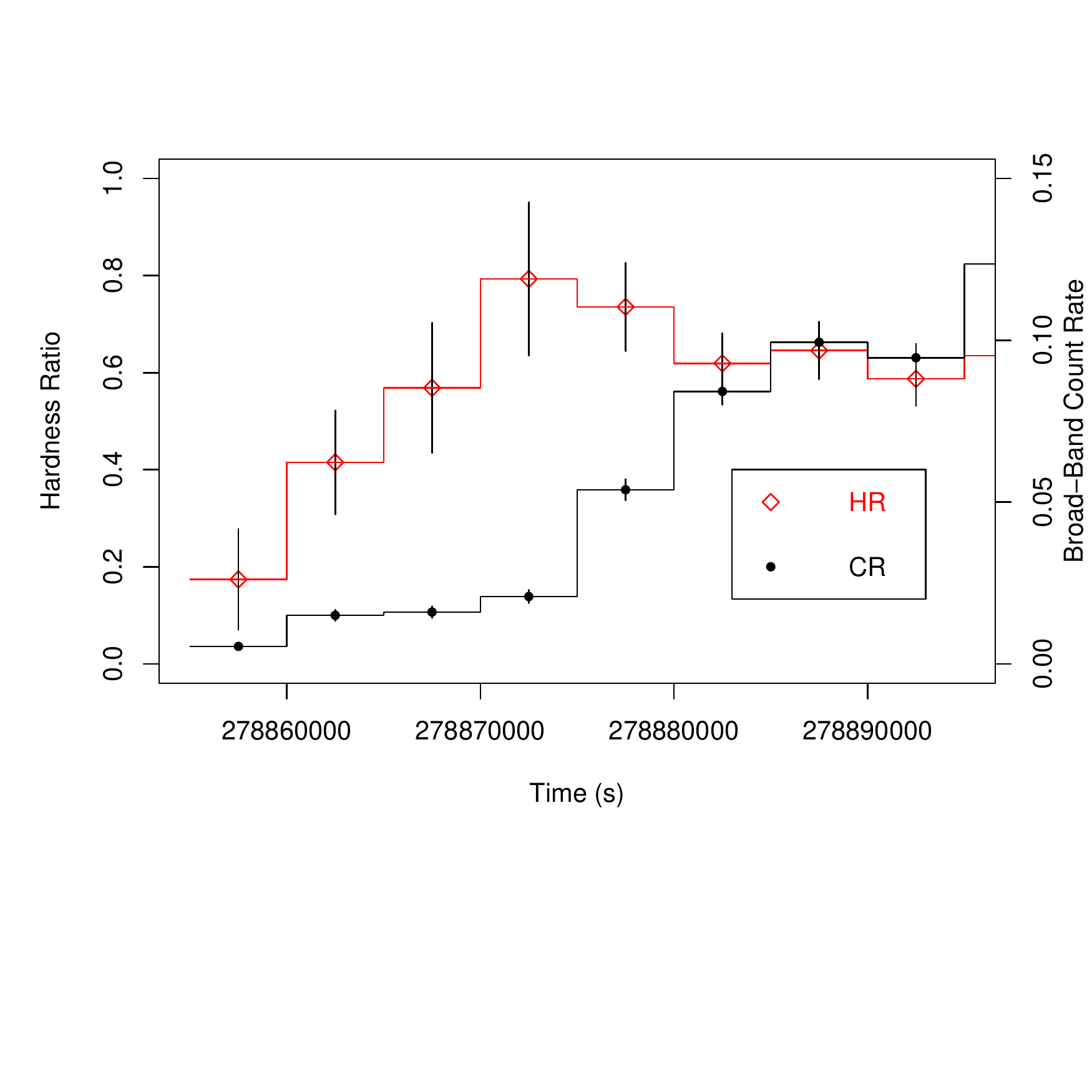}
\caption{{\bf  Plot of hardness ratio during the eclipse egress seen in observation $07082$, binned at 5000 sec intervals to reveal the steady rise in HR as the BH moves out of eclipse. The flux increase appears to lag the rise in hardness ratio. The midpoint of this eclipse occurred on MJD 54040.87, which defines the reference epoch used through out this article.}}
\label{fig:HR_profile}
\end{center}
\end{figure}

The correlation between HR and CR was investigated by plotting the X-ray CMD (CR vs HR) in Figure~\ref{fig:chandra_CMD}. This reveals a split between high and low flux states, which occurs at a count rate of 0.06 c/s, corresponding to a luminosity of 4$\times$10$^{37}$ \lx (assuming $D_{IC10}$=600 kpc, and a power-law spectrum with $\Gamma$=1.5, $N_H$=5$\times$10$^{21}$). This CR also happens to divide the eclipse and un-eclipsed CR values seen in observations 2 \& 3. Interestingly, observations 4, 8 \& 9 all lie completely below the 0.06 c/s threshold. The full  duration of X1's eclipses is about 7 hrs, with 5 hrs spent at minimum flux. The monitoring observations are all $<$15 ksec (a similar duration to the eclipses) so  it is possible that the 3 low flux observations are all ``in eclipse", since the odds of a single observation falling mid-eclipse is about 14\% (5/35). During the eclipse egress of $ObsID$ 07082 (Figure~\ref{fig:rate_and_hr_all} panel 3) the HR is seen to rise steadily, Figure~\ref{fig:HR_profile} shows the events binned at 5000 s intervals to improve the S/N, demonstrating that the eclipse profile is energy dependent. Curiously the rise in hardness ratio {\it precedes} the corresponding rise in flux that signals egress. 

\section{Refining the Orbital Period}
\label{periodsearch}
Multiple light-curve measurements spread over many orbital cycles can enable very high precision period determination. The key lies in being able to unambiguously line-up stable features observed in widely separated observations. This process of phase connection can only be achieved if one already has a well constrained measurement of the period, and the spacing between observations is such that the phase error does not accumulate to the point that the absolute phase of known features becomes ambiguous. This approach has been widely used (throughout the history of astronomy) to determine the orbital periods of binary stars (e.g. \citealt{Singh2002} in the case of Cyg X-3). With a sufficiently accurate period we can identify the orbital phase of the X-ray observations, and existing optical data-points. Very high precision measurements of the orbital period are required in order to detect period derivatives in HMXBs; model predictions for IC 10 X-1 suggest $\Delta P/P \leq 10^{-6}$ \citep{Tutukov2013}.

In the case of the \chandra/\xmm ~monitoring of IC10 X-1, we have 12 observations (Table~\ref{tab:dataset}) one pair of which (07082, 08458) show unambiguous eclipse egress and ingress respectively \citep{Prestwich2007}. The remaining 10 observations are randomly spaced, and so should contain a range of orbital phases. The final \xmm ~observation is special because it covers a full orbital cycle and thus provides a chance to confirm the picture arrived at from the \chandra ~monitoring.  It also greatly extends the baseline, such that a 1 second departure in the ephemeris becomes a $\Delta \phi =  2\% $ phase shift over the 2500 orbital cycles encompassed by a decade. In order to interpret these data, we need to know the orbital phase of each one. 

The orbital period of IC10 X-1 was first  determined from a dense week-long series of Swift X-ray observations by \cite{Prestwich2007}, yielding P= 34.40$\pm$0.83 hr (123840$\pm$2988 s). Shortly thereafter an optical radial-velocity study by \cite{SF} using Keck produced a more precise ephemeris P= 34.93$\pm$0.04 hr  (125748$\pm$144 s).  Maintaining phase connectivity over many orbital cycles requires that the binary period be known to a corresponding level of precision. The published period has a precision of $\Delta P/P$=1.145$\times$10$^{-3}$ (approx. 0.11\%). Accordingly the accumulating orbital phase error exceeds 10\% after 88 binary periods (128 days) have elapsed, and total loss of phase ($\Delta \Phi$=0.5) occurs after 437 cycles, or 1$\frac{3}{4}$ years.  The largest spacing between X-ray observations is $\sim$ 3 years and the smallest is 2 days (See Table~\ref{tab:dataset}). Thus in principle phase connection can be achieved provided that recognizable features can be discerned in the light-curve segments. 

We explored a set of trial periods centred on the published orbital period of \cite{SF} and extending by $\pm$1\% on either side. Folded light-curves were generated for 1500 trial periods in 1 second increments to create a movie of the resulting phase-folded light-curve.  Visualized in this way it was possible to examine how the various light-curve segments fit together, and see the effects of changing the period and epoch.  Naturally the more distant an observation is from the epoch, the more its phase moves in response to a very small change in period. Each event's arrival time was transformed to orbital phase using the {\it ObsID 07082} mid-eclipse time to define phase $\Phi$=0.5, which set the the folding zero point or epoch ($T_0$ = 278801348 mission seconds = MJD 54040.87). To avoid negative phases we subtracted 100,000 cycles of the trial period from the actual time of mid-eclipse, as given in Equation~\ref{eqn:phase}. The \xmm ~data were included by folding the {\it XSA} pipeline light-curve on the same ephemeris. The time systems used to tag \chandra ~and \xmm ~data share the same reference point (elapsed seconds since 1998.0 TT) so no time conversion is required.

\begin{equation}
\Phi = \frac{(t - T_0 - 100000P)}{P}
\label{eqn:phase}
\end{equation}
 
\begin{figure*}
\begin{center}
\includegraphics[angle=0,width=14cm]{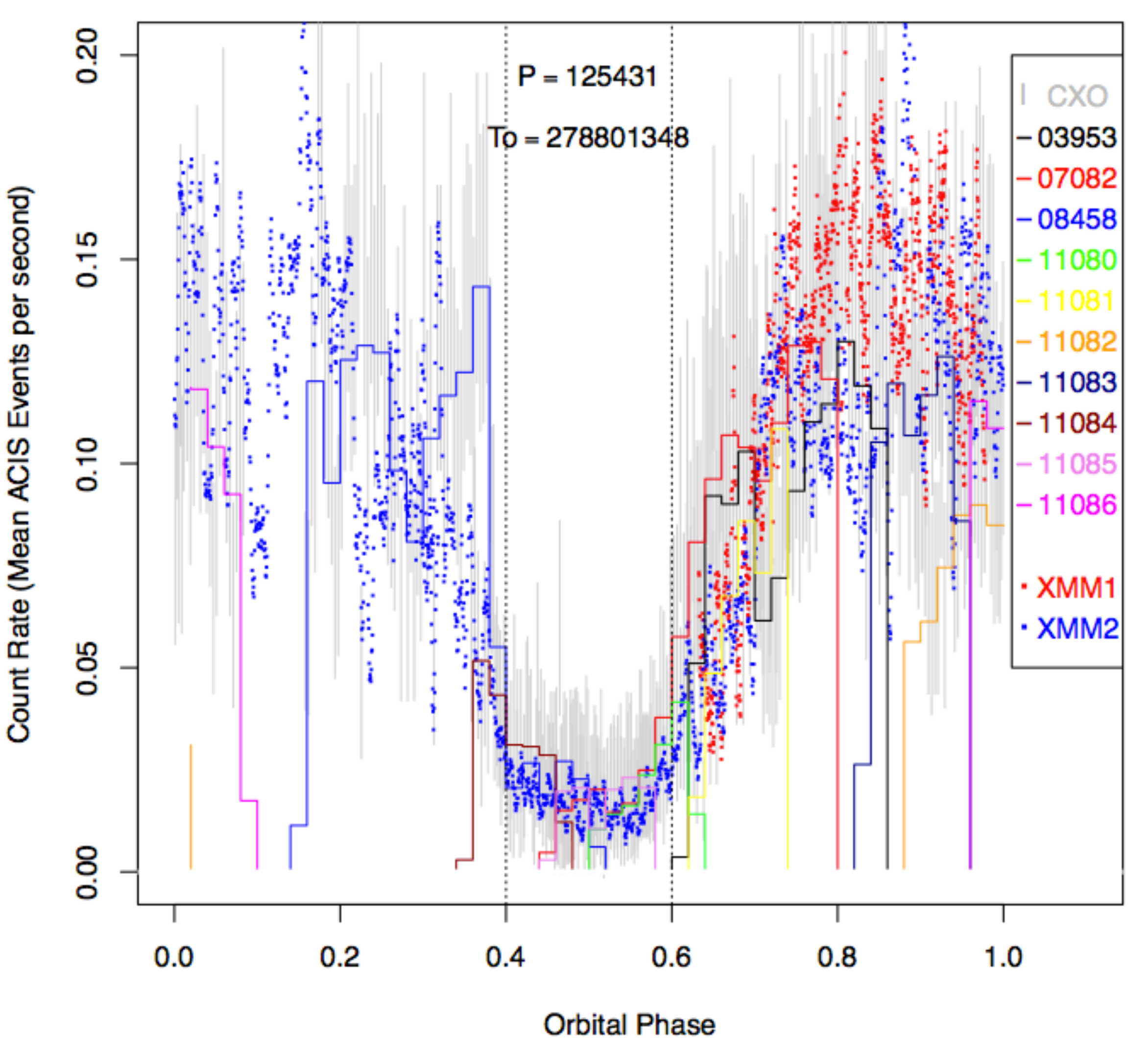}
\caption{{\bf  Phase-folded light-curve for all 10 Chandra and 2 \xmm ~observations, for a trial period of P = 125431 s (1.45175 d) , based on visual inspection and eclipse-flux analysis of the light-curves produced by a range of trial periods spanning 1\% either side of the published orbital period. This period lies within 2$\sigma$ of the published value from the optical RV study of \cite{SF}. An animation of the folded light-curve segments stepping through 1500 trial periods is available in the online journal. }}
\label{fig:eclipse}
\end{center}
\end{figure*}

Once all of the events (or light curve bins) are transformed to phase, they are grouped into histogram bins within each contiguous observation. This yields a set of 10 histograms, the binning scheme (breakpoints between bins) is chosen so that each dataset is binned to an appropriate resolution given the number of events present and the duration of the observation. We tried a variety of algorithms including fixed width (0.1 or 0.05, and 0.02 in phase) and the \cite{scott}  formula for optimal bin choice. A change of 1 second in the trial period represents a cumulative phase shift of 0.002 per year, accumulating over a decade to 0.02. Consequently the smallest useful bin-size is of order $P/100$ or so. The count rate in each phase histogram bin is calculated as the number of events in the bin, divided by its width in seconds. The count rates for each observation are then divided by the normalized exposure-map value for the extraction region in the original \chandra ~images. No correction is applied for extraction region size, since we used regions scaled to a fixed fraction of encircled energy (determined from the HRMA PSF model). The phase histogram of all 10 observations are plotted in different colours in Figure~\ref{fig:eclipse} which appears in the online journal as an animation stepping through the range of trial periods; the figure shows a single frame at the ``best period''.  As a comparison, we also folded a binned light-curve (100 sec time bins, and incorporating exposure map correction) for each trial ephemeris, and plotted the resulting points and their statistical error without re-binning by phase. This representation shows the intrinsic scatter in the flux values, and is shown in grey in Figure~\ref{fig:eclipse}. 

\begin{figure}
\begin{center}
\includegraphics[angle=0,width=9cm]{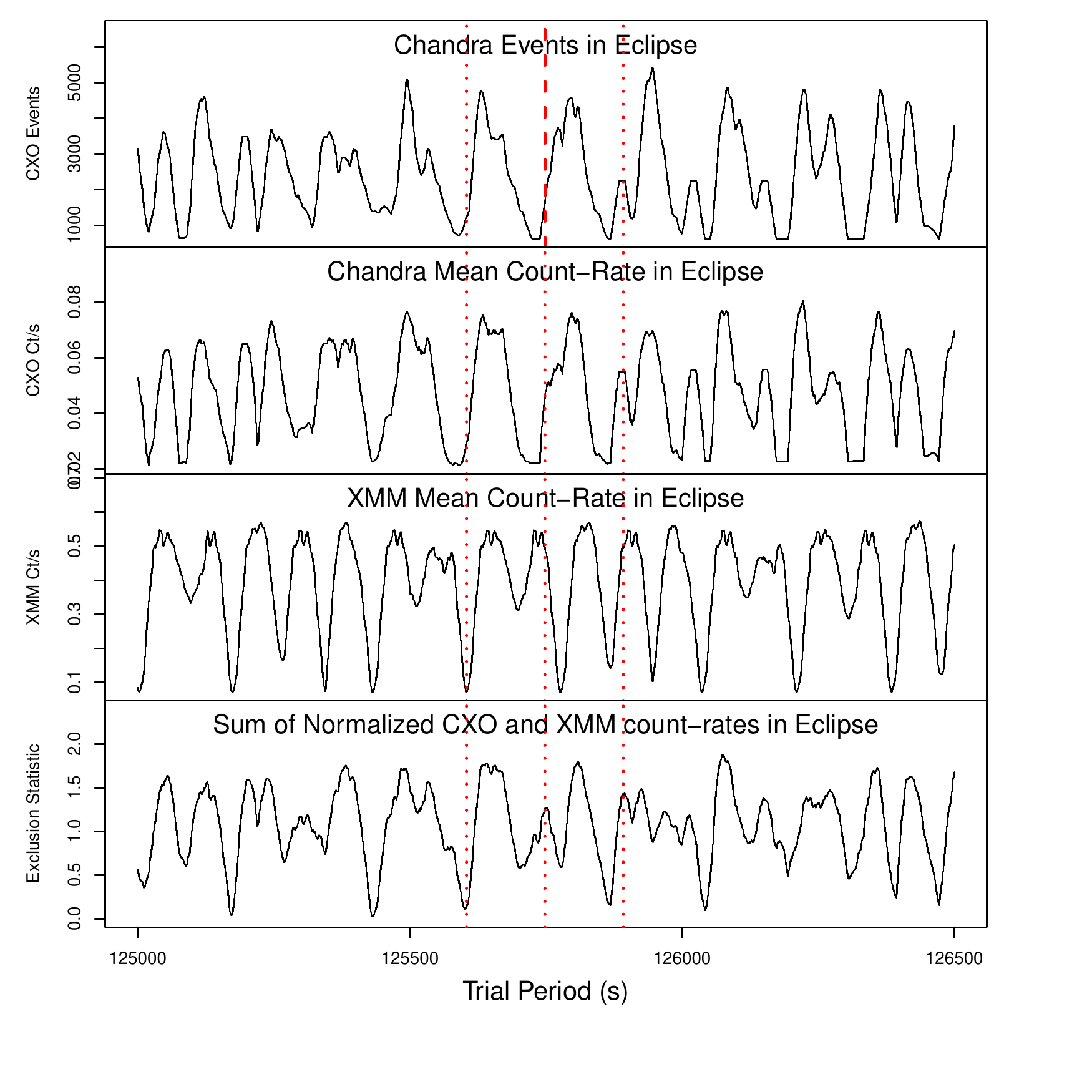}
\caption{{\bf Eclipse Periodograms created by the \chandra ~and \xmm ~events and count rate inside the reference eclipse (phase=0.4-0.6) for trial periods spanning 1\% either side of the established orbital period. Vertical dotted lines indicate the \cite{SF} period and its 1$\sigma$ uncertainty.  Minima correspond to trial periods that produce a consistent ephemeris. The \chandra ~and  \xmm ~data each define an independent set of minima.The bottom panel shows the intersection of these sets, resulting in a smaller set of mutually consistent periods. }}
\label{fig:trialperiods}
\end{center}
\end{figure}

Visual inspection showed no trial period places all low-flux observations in eclipse and simultaneously all high flux observations outside of eclipse, leading to three possible scenarios : (1) either the X-ray source can sometimes become visible during eclipse (requiring it to become more spatially extended). (2) The system alternates between different states such that it can show low flux even when not eclipsed. (3) The true period lies slightly outside the reported range.   

Applying the condition that no high-flux values should conflict with the eclipse profile defined by the near-contiguous observations ($ObsIDs$ 07082, 08458), but allowing low-flux values to fall outside of the eclipse, we devised a numerical test to search for candidate periods. It was not possible to use other established methods (e.g.  O-C) because most of our samples of the light-curve contain no {\it a priori} recognizable features.  At each trial period, we recorded the number of photon events falling inside the eclipse, expecting that the correct period will yield a minimum value. By treating the \chandra ~and \xmm ~data separately we have two independent test statistics.  We first define a reference eclipse and epoch, and set a nominal eclipse width as follows. 

We chose the folding zeropoint ($T_0$= 278801348s = MJD 54040.87) such that the midpoint of the \chandra ~``reference eclipse" lies at phase 0.5 in our plots. The full width of the reference eclipse is 0.2 in phase, so the region $\Phi$= 0.4-0.6 should contain minimal flux for an acceptable binary period.  At each trial period, we record (separately) the number of ACIS and \xmm-PN counts falling inside the eclipse, and then plot the results in Figure~\ref{fig:trialperiods}.   Note that two slightly different versions of the Chanda statistic are plotted: one from the un-binned event times, and the other from the binned light-curve (which includes exposure map correction). The value of these statistics only changes with period if the number of events falling within the selected eclipse range changes, so there are some flat regions in the graph. Figure~\ref{fig:trialperiods} reveals a series of local minima roughly evenly spaced in trial period-space. Each minimum is very narrow, and we can definitively exclude any trial period not lying in such a minimum.  The resulting set of candidate periods can then be compared with other independent constraints in order to see if they intersect.

\begin{figure}
\begin{center}
\includegraphics[trim= 0 0.5cm 0 2cm, clip,angle=0,width=9cm]{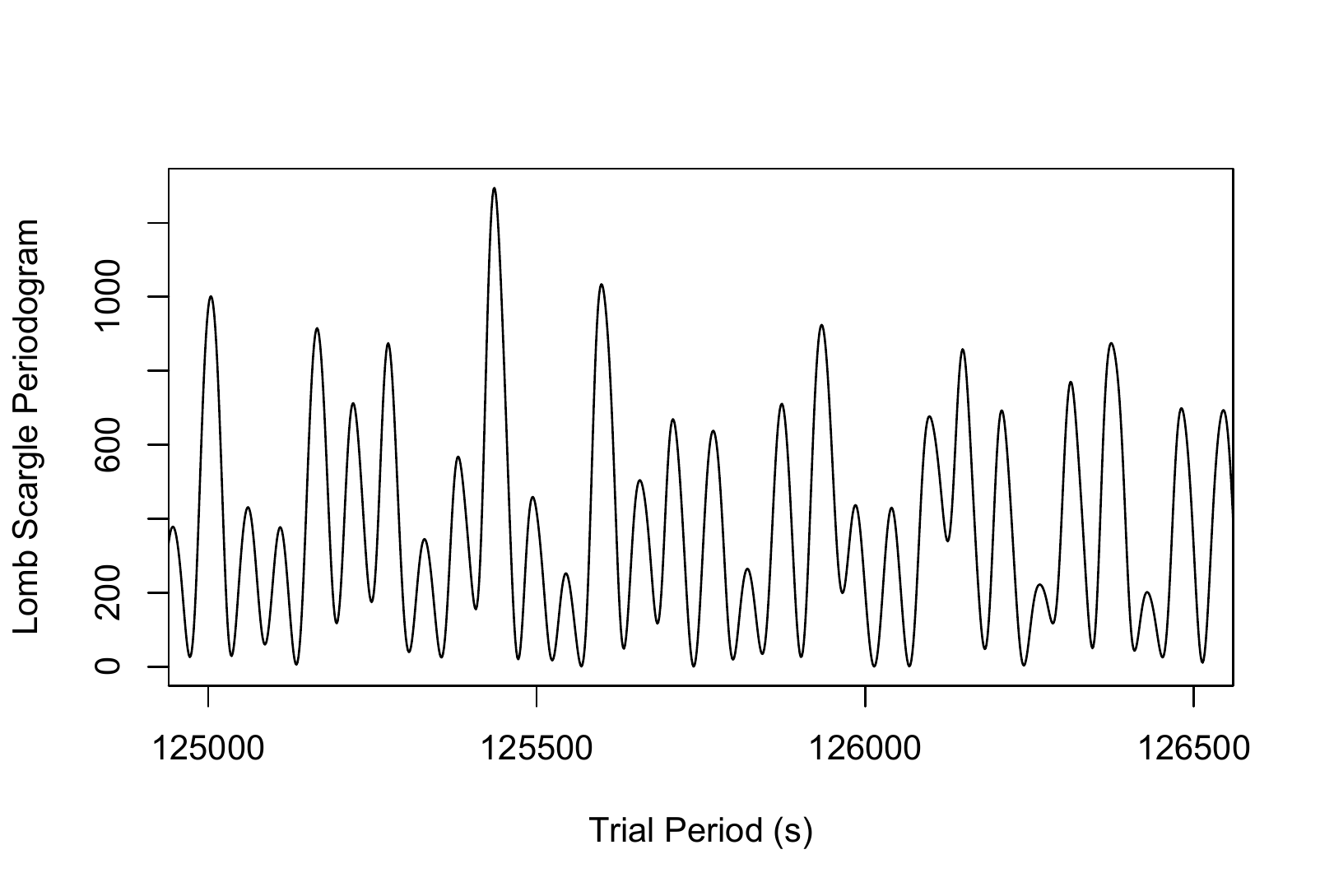}
\caption{{\bf  Lomb-Scargle Periodogram for the combined \chandra + \xmm ~light-curve. The highest peak is at a period of 125435s or 34.84306 hr.  
Monte-carlo simulations reveal that the pattern of additional peaks is consistent with aliasing. (The 99.9\% significance level is 8 for an un-correlated light-curve.)}}
\label{fig:scargle}
\end{center}
\end{figure}

Within the 3$\sigma$ period range reported by \cite{SF}, indicated in Figure~\ref{fig:trialperiods},  we find 8 minima in the \chandra ~statistic, and 7 in the \xmm ~statistic.  Most of these minima are mutually exclusive between the \xmm ~and \chandra ~series, leaving just 4 candidate periods. The minima have a regular spacing driven by the difference in period needed to accumulate phase shift in integer multiples of the orbital period for the various observations comprising the time series. Visual inspection of the corresponding folded light-curves reveals that  one of the candidate periods (P = 125431s = 1.45175d) places more of the low flux observations inside eclipse, as shown in Figure~\ref{fig:eclipse}.  The ``best" period simultaneously places the \xmm ~observations of 2003 and 2012 right on top of the \chandra ~reference eclipse (2006) and reveals close consistency between the eclipse profiles in the two \xmm ~observations. The rising portion of the 2003 observation is thus confirmed to be an eclipse egress, as its profile is an exact match to the 2012 egress within errors.   Evidently the source luminosity and extent of the emission region is the same in both observations. Flaring of similar amplitude and duration is also seen at both epochs. 

To estimate the uncertainty of the orbital period measurement, we again consider the accumulating phase drift across the entire dataset. Relative to the vertical lines at $\phi$ = 0.4, 0.6 plotted in Figure~\ref{fig:eclipse} neighboring trial periods at $\pm 1 s$ cause features to shift by $\Delta \phi =  0.02$ such that a drift of 2 seconds in either direction causes the eclipse feature to unravel. 

As a further check on our candidate period, we computed a Lomb-Scargle periodogram (LSP) for the entire Chanda light-curve binned at 100 second resolution. For \chandra ~data alone the resulting highest peak is at P= 125439s. We then recomputed the LSP for the combined \xmm ~and \chandra ~time series which is plotted in Figure~\ref{fig:scargle}, obtaining P =  125435 s (Half-Width at Half-Maximum = 20 s). To assess significance, uncertainty and the influence of the observing cadence we ran a 1000 iteration monte-carlo, randomizing the light-curve points among the actual observation times and re-computing the LSP. We find all features in Figure~\ref{fig:scargle} lie far above the power distribution for un-correlated light-curve; the observed pattern in the LSP is therefore due to aliasing. We performed a second monte-carlo involving the addition of a periodic component matched to the observed modulation (Period = 125435s, Amplitude = 0.05 c/s) to the randomized flux values. The phase of the simulated periodicity was changed at random for each trial. The recovered periods ranged between 125432 - 125438 with a standard deviation of 1.1 s. A similar pattern of alias peaks occurred in all trials and never exceeded the power at the true period. The eclipse folding analysis and the LSP are therefore consistent, having different systematic errors.


\section{Analysis of the X-ray Eclipse Profile }

For a point-like X-ray source and a primary star with a clearly delineated boundary, a total eclipse should appear as a nearly rectangular dip. The duration of the minimum gives the diameter of the larger star, and the duration of the ingress or egress gives the diameter of the X-ray emission region. In reality more complex morphology results from X-ray scattering in the wind of the companion, the existence of extended emission and absorption components (e.g.  accretion disk, corona, impact hotspot), partial eclipses, and gravitational distortion of the companion \citep{Orosz2007}. The following processes can in principle blur the edges of the eclipse: (a) In the typical low count-rate regime the binning process needed to visualize the light-curve will automatically impose a loss of timing resolution. (b) The X-ray source may indeed be extended (In this context extended means its diameter is larger than the product of the timing resolution and the relative velocity between the two stars.), (c) The column density of stellar wind particles increases rapidly as the BH approaches first contact so increasing absorption could blur the otherwise sharp cutoff. (d) reprocessed X-rays scattering through the wind of the primary star could continue to be seen after the X-ray source passes second contact.  (e) Finally, intrinsic X-ray variability is convolved with the underlying eclipse profile.

\begin{figure}
\begin{center}
\includegraphics[trim= 0 1cm 0 0, clip,angle=0,width=9cm]{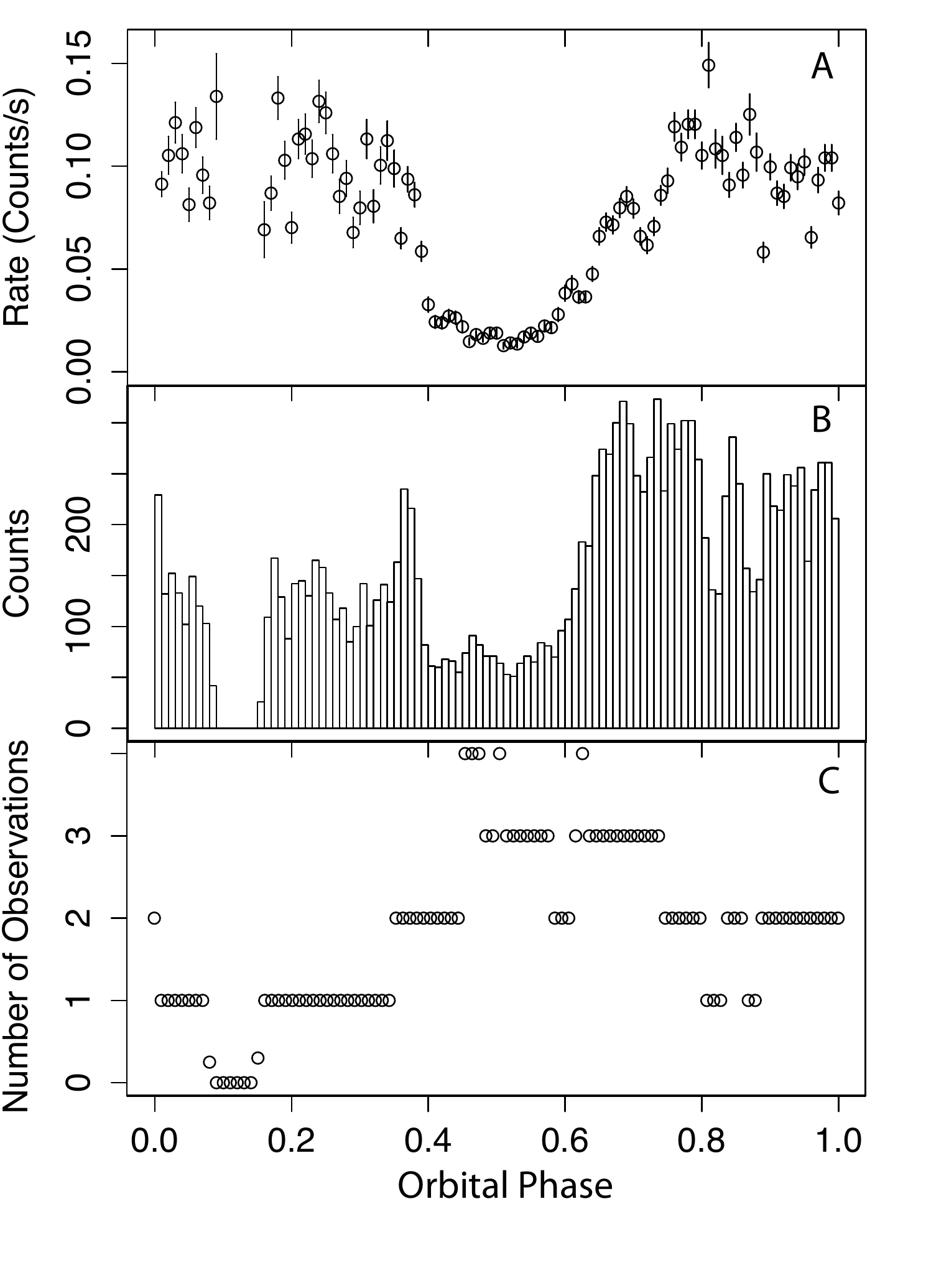}
\caption{{\bf  The average Chandra eclipse profile (A) was computed by dividing the total number of counts in each Phase bin (B) by the number of observations at each phase bin (C) to effect a correction for fractional exposure.}}
\label{fig:cxobinned}
\end{center}
\end{figure}

The eclipse profile seen in the \chandra ~data (Figure~\ref{fig:eclipse}) is deep and well defined. The duration of minimum flux is approximately 5 hours (0.15 in phase), during which the background subtracted count rate never reaches zero, remaining above the rate observed in the diffuse emission surrounding the point source. The full-width of the X-ray eclipse is about 7 hours $\Phi$ = 0.2$\pm$0.05. The region is marked in Figure~\ref{fig:eclipse} by vertical dotted lines. To study the shape further we generated an average eclipse profile from the entire \chandra ~dataset to improve S/N, and correct for fractional exposure. The resulting average 0.3-8 keV profile is shown in Panel A of Figure~\ref{fig:cxobinned}. The raw folded/binned light curve and the number of observations contributing to each phase-bin are shown in Panels B \& C respectively. 

The \chandra ~eclipse profile shows an asymmetry between the ingress and egress. The flux undergoes a very steep drop from phase 0.38 - 0.42, remains at minimum until phase 0.6 and then climbs steadily until phase 0.7.  Many other accreting binaries (HMXBs, LMXBs, and CVs) show a comparable asymmetric eclipse shape which is regarded as the signature of an accretion-disk hotspot. As the compact object moves into eclipse it is partly obscured by matter streaming onto the accretion disk, because of angular momentum conservation this hotspot trails the BH in its orbit, when the BH re-emerges it leads the hotspot, emerging clear of the accretion stream while the hotspot is still hidden, leading to a deficit in intensity at egress (See e.g. ~\citealt{Rutten1992}).  We also see higher levels of variability before ingress than at other times, similar to the BH-HMXB M33 X-7 \citep{Pietsch2006,Orosz2007}.

The eclipse exhibits ingress and egress stages of duration $\phi \sim 0.05$, indicating a length scale of 1.5-2 \rsun for scattering and/or extended emission. There are several independent observations of eclipse egress (two from \xmm  ~and panel C of Figure~\ref{fig:cxobinned} shows the coverage by \chandra), all show a feature at phase 0.7. Either it is a dip centred at phase 0.7, or a brightening just prior which ends at phase 0.7. This feature is evident in binned average light-curve shown in panel A of Figure~\ref{fig:cxobinned} constructed from all the \chandra ~data. If it truly is stable, then it could be due to the the emergence of the trailing side of the accretion-disk with its attendant hotspot.

The eclipse is broader and more gradual in the \xmm ~data than in the \chandra ~data. We attribute this to energy dependence of the eclipse profile originating from absorption and scattering by the stellar wind. \xmm ~has a greater effective area at low energies resulting in it seeing a broader eclipse.   

\begin{figure*}
\begin{center}
\includegraphics[angle=0,width=18cm]{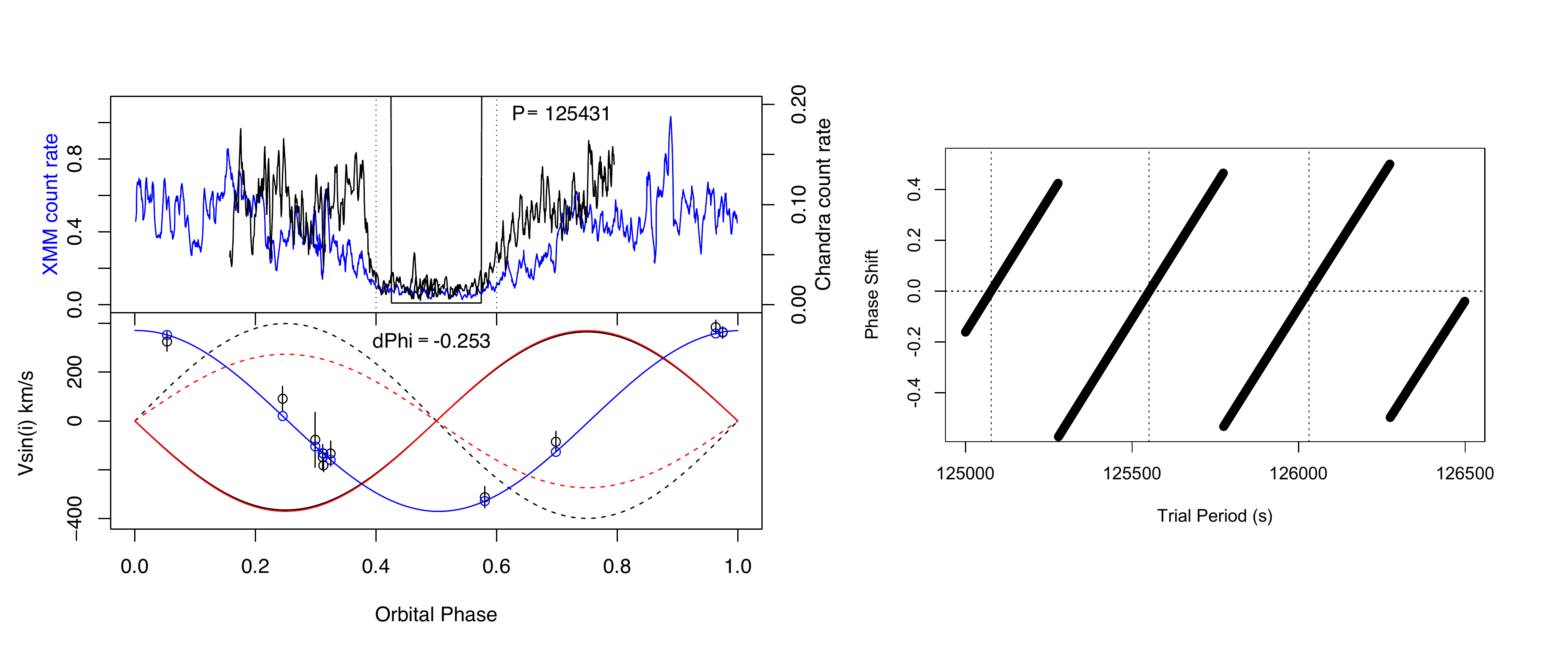}
\caption{{\bf Comparison of X-ray Eclipse Ephemeris with Optical Radial Velocity Data.  Left Panel: Reference Eclipse (top) and radial velocity curves (bottom) using the ephemeris (P = 125431s or 1.45175d , $T_0$ = MJD 54040.87) and assuming $i=90$. The solid red line is the model curve for the WR star, whose motion changes sign from approaching to receding (i.e. inferior conjunction) at the midpoint of the reference eclipse. Dashed lines are the model RV curves for the BH that would result for the two mass-scenarios considered in the text. Points with error-bars are the published He II RV values of \cite{SF} folded on our ephemeris. Solid blue curve is the best-fitting sinewave to the RV points, used to determine the phase shift between the RV points and the model. 
Right Panel: The Phase Shifts resulting from repeating this analysis over a range of trial orbital periods. Dotted lines indicate the specific trial periods for which the RV derived inferior conjunction occurs at Phase=0.5}}
\label{fig:EclipseModel}
\end{center}
\end{figure*}

\subsection{Calculating the Binary Parameters and the Size of the X-ray Shadow}
\label{sect:orbitcalc}

Following the usual procedure for determining the dimensions of an eclipsing binary system (e.g. \citealt{CS}, \citealt{CO}) we assume circular orbits and inclination i=90\degree, and make use of the following results from \cite{Clark2004}, \cite{Prestwich2007}, and  \cite{SF}:  The radial velocity semi-amplitude of the Wolf-Rayet star  is 370$\pm$20 km/s; the mass of the WR is most likely 35 \msun, requiring the mass of the BH to be 32$\pm$2.6 \msun. The lowest acceptable mass for the system has $M_{WR}$=17 and $M_{BH}$=23.1$\pm$2.1 \msun. In the former case the WR star is the most massive member of the binary, and in the latter the BH is the more massive. These alternative scenarios are summarized in the first 4 lines of Table~\ref{tab:BHHMXBs} of the Appendix. We estimate the  duration of minimum X-ray flux in Figure~\ref{fig:eclipse} to be 5 hours, with ingress/egress each lasting about 1 hour.

For circular orbits, the following relations follow from newtonian mechanics: $m_1/m_2$ = $v_2/v_1$ = $a_2/a_1$ and in addition the tangential velocities ($v_1$, $v_2$) and separations ($a_1$,$a_2$) are constant and are related via Kepler's 3rd Law.

If the orbital plane lies along the line of sight, (which it must be close to since the system has deep eclipses), then the relative velocity between the two stars is $v_{rel}$=$v_1$+$v_2$, and as seen from the earth this is equal to the distance travelled by the BH relative to the WR star divided by the eclipse duration ($v_{rel}=2R_{WR}/T_{eclipse}$). Hence the measurement of the eclipse duration yields $R_{WR}sin(i)=T_{eclipse}(v_1+v_2)/2$. 
    
Using these relations, we calculate the range of radii corresponding to the `most massive' and `least massive' cases to be $R_{WR}^{max}$ = 10.1 \rsun and $R_{WR}^{min}$ = 8.4 \rsun.
The respective orbital separations are ($a_1+a_2)sin(i)^{max}$ = 1.527$\times10^{10}$ m (21.923 \rsun) and ($a_1+a_2)sin(i)^{min}$ = 1.286$\times10^{10}$ m (18.46 \rsun). The inferred separation depends weakly on the masses and inclination (given the observational constraints on the mass function and inclination). The results are given in Table~\ref{tab:BHHMXBs} and compared to the same calculations for  other known BH-HMXBs as discussed in Section~\ref{sect:conclusions}. The corresponding estimates for the length-scale of the wind-scattering region and/or extended X-ray emission (as inferred from the duration of ingress/egress portions of the light curve) are 2.0 \rsun and 1.7 \rsun.  

We performed identical calculations for the other known BH-HMXBs and summarize the results (and citations) for reference in the Appendix and Table~\ref{tab:BHHMXBs}. We notice that the other WR systems (NGC 300 X-1, Cyg X-3, and the object in NGC 4490) show similarly broad X-ray modulation. Of these only NGC 300 X-1 is known to be eclipsing, and our calculation shows that it too has eclipse shadows of order 10 \rsun. In contrast the SG system M33 X-7 has an eclipse duration in good agreement with its stellar radius as inferred from other means, as is typically seen in eclipsing neutron-star + SG HMXBs.  

With these approximate dimensions we can estimate the possible range of inclination angles for the system such that a point like BH undergoes total eclipses (while the extended or scattered  emission could remain visible). The angular radius of the star (or its dense wind) as seen from the BH is $sin (\theta) \sim R/(a_1+a_2)$ For the radii calculated above, this results in $\theta$ =  27\degree ~for either scenario (both the separation and stellar radius  increase with mass). The minimum inclination angle for eclipses is then $i$ = 90 - $\theta$, which gives $i$ $>$  63\degree. We also calculated the RV semi-amplitude that would result for different values of the inclination, while keeping $P_{orbit}$ fixed and remaining within the allowed range of masses. We found that RV$_{max}$ remains below the maximum possible 390 km/s (RV+error) for $i$  $>$ 71\degree. For the WR star itself to eclipse the BH the limiting inclination is $i<85$\degree.

\section{Comparison between X-ray and Optical Radial Velocity Observations}
\label{sect:RV}
As a direct comparison between our X-ray eclipse ephemeris and existing optical spectroscopy, we analyzed the published He II [{\small $\lambda\lambda$}4686]  line radial velocity data of \cite{SF}. 
We used a similar procedure to the one described in section \ref{periodsearch} to test trial periods against the reference eclipse.  At each trial period we transformed the optical observation times (published as HJD and converted to seconds since 1998.0 using {\it xtime}\footnote{http://heasarc.gsfc.nasa.gov/cgi-bin/Tools/xTime/xTime.pl}) to X-ray eclipse phase (where 0.5 is defined as mid-eclipse as in Equation~\ref{eqn:phase}  and Figures~\ref{fig:eclipse} \& ~\ref{fig:cxobinned}) and then fit the RV data with an anti-sinewave of amplitude 370 km $s^{-1}$. The only free parameter in the fit is a phase-shift. The value of the phase-shift quantifies agreement between the RV data and the X-ray ephemeris. Our method tests the assumption that the mid-point of the reference eclipse corresponds to inferior conjunction (the point in the orbit when the WR star switches from moving towards us and begins moving away).  The resulting array of phase-shifts is plotted against trial period in Figure~\ref{fig:EclipseModel}  revealing just 3 trial periods (125077 s,  125551 s, 126031s) that satisfy this criterion.  The phase shift between the X-ray and optical observations should be consistent with zero if the HeII line originates on the WR star, while values close to $\pm$0.5 would place the emission close to the black hole. Of the 3 periods thus identified only P=126031 falls in a minimum of the X-ray eclipse periodograms plotted in Figure~\ref{fig:trialperiods},  both of the others being ruled out. The lone surviving period does not give the deepest eclipse minimum, and is not the period identified in the Lomb-Scargle periodogram. Examination of the eclipse-folding animation (accompanying online data for Figure~\ref{fig:eclipse}) shows this period is problematic as it places all of the low-flux Chandra observations far outside eclipse.

Our preferred X-ray period of 125431s (1.45175d) produces a phase shift of -0.25 (-$\pi$/2 radians) which {\it either} places the He II emitting material trailing the WR star around its orbit by up to 90\degree or implies that the eclipse occurs substantially before inferior conjunction.  If the He II line originates in the wind, it raises the question of the origin of its asymmetry with respect to the star. Photoionization of the wind by X-rays will restrict the physical conditions for line-emission to regions shadowed by the star and regions sheltered by sufficiently high column-density of wind particles. Depending on the exact geometry, the observed RV curve will involve contributions from orbital motion, wind velocity, and rotational velocity of the star. 

An appealing mechanism exists for producing an asymmetric density structure in the binary system, which naturally results in eclipses that precede inferior conjunction. Persistent spiral shock-structures known as pinwheels or dust spirals form in the winds of close binaries containing WR stars. The prototype colliding-wind pinwheel is WR 104  \citep{Tuthill2008,Monnier2011} consisting of a WR + OB in a 241.5 day orbit. The high velocity wind of the WR collides with the weaker wind of the secondary forming a shock front. Behind this front the WR wind stalls, pushing up its density and allowing dust grains to form. In the case of WR 104 the region of elevated density trails behind the O-star while moving outwards, tracing out a spiral. If the same physics occurs in IC 10 X-1 then the dense region is trailing behind the BH and will lie between the observer and the BH when the BH is moving directly away from us (and hence the WR moving directly toward us). The deepest part of the X-ray eclipse should coincide with the time of maximum blueshift of the He II line. This situation is what we see in Figure~\ref{fig:EclipseModel}. If this interpretation is correct the size of the dense region can be approximately inferred from the eclipse duration and the circular orbital velocity of the BH (See Table~\ref{tab:BHHMXBs}), resulting in 5-7 $\times 10^{6}$ km. Equivalently during the $\sim$5 hr eclipse, the BH moves through 5/35 of its orbit. The wind velocity encountered by the BH will be lower in IC 10  X-1 than in WR 104 due to the smaller orbital separation. The orbital (and angular) velocities however are much higher, consequently rotation-induced effects need to be incorporated into any model.  The phase shift can also relieve the geometric constraint on the orbital inclination. If X-ray eclipse does not correspond to inferior conjunction then the WR star is not actually eclipsing the BH, and the X-ray modulation could  be entirely due to absorption and scattering in the wind.

\section{Spectral Changes}

ACIS-S pulse-height spectra with sufficient counts for spectral analysis are available for all 10 \chandra ~observations (Table~\ref{tab:dataset}). 
Using the 2003 ACIS-S data \cite{BB} found that the point source is embedded in spatially extended X-ray emission, and described a two component spectral model consisting of a power-law ($\Gamma=1.8$) plus a thermal plasma ($kT=1.5$ keV), accompanied by an absorbing column consistent with IC 10 ($N_H=6 \times 10^{21} cm^{-2}$). The power-law component dominates the point-source spectrum, and the thermal component appears to be associated with the extended emission.   \cite{Wang2005} found the 2003 \xmm ~PN+MOS spectrum (which has higher S/N) is more consistent with a comptonized multicolour disk blackbody (MCD) with inner temperature $T_{in}$=1.1 keV, and that the \chandra ~data are well described by the same model. The parameters of the MCD model are such that a high value of the black hole's spin is required in order to reconcile the spectral fit with the large mass implied by the mass function. Residual structures reported by \cite{BB} as evidence of photoionization lines were interpreted by \cite{Wang2005} as artifacts of the data reduction which was complicated by pile-up correction and the presence of background flares.

\begin{figure}
\begin{center}
\includegraphics[angle=0,width=9cm]{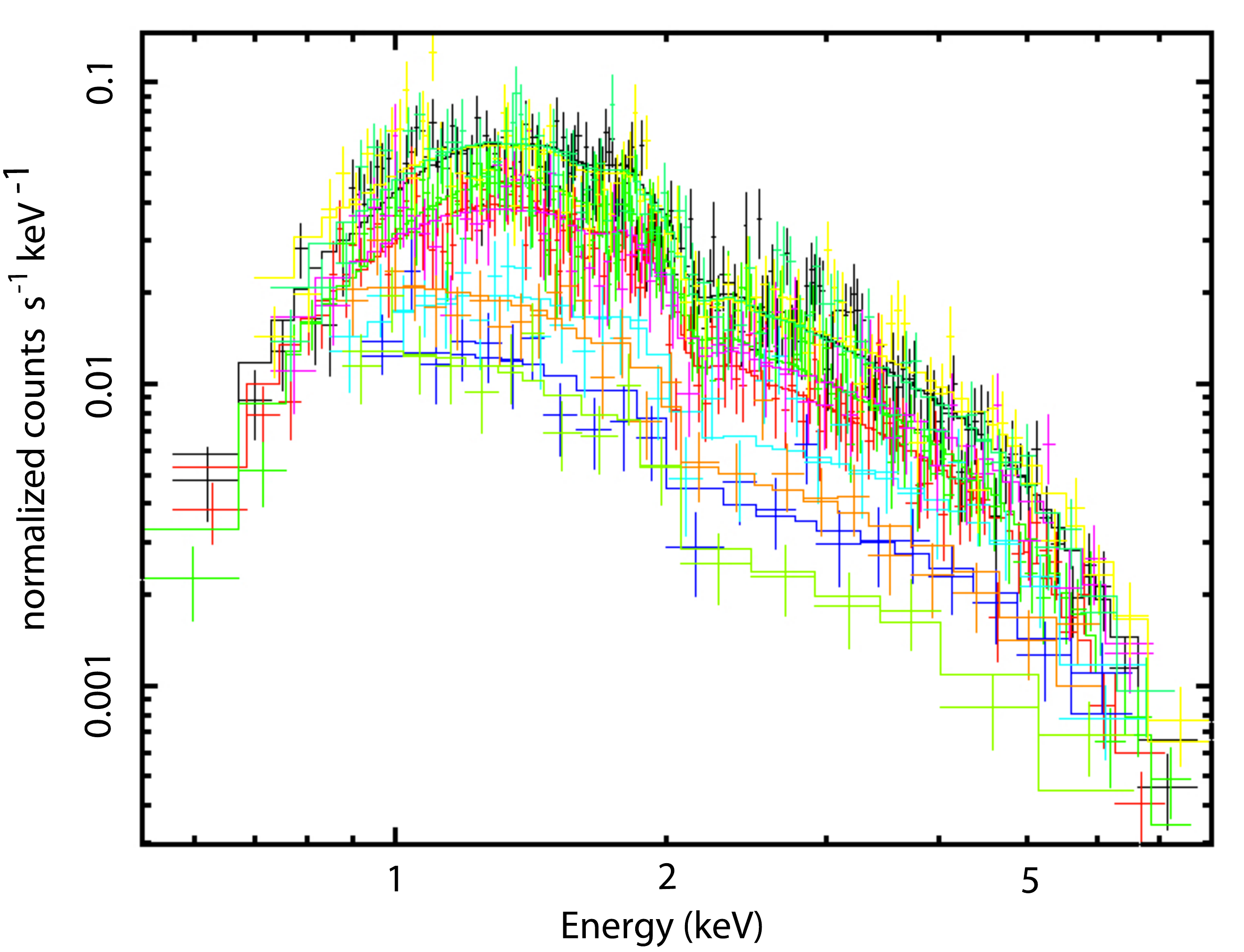}
\caption{{\bf  Chandra X-ray spectra of the 10 observations in Table~\ref{tab:dataset}. Each spectrum is fit by an absorbed power-law, the parameters are given in Table~\ref{tab:xspectra} (The spectra can be identified from their flux).  }}
\label{fig:multispec}
\end{center}
\end{figure}

In order to search for spectral variations the 10 ACIS-S pulse-height spectra are plotted on the same axes in Figure~\ref{fig:multispec}, and the results of the spectral fitting to each observation are given in Table~\ref{tab:xspectra}. For consistency (in the face of varying S/N) we used the simple absorbed power-law model. Our fitted model parameters for $ObsID$ 03953  are consistent with those found by \cite{BB}  for the same observation, although we did not include the thermal component, and made no correction for pileup (which is mild at these count rates). Among observations there are no clear trends in spectral index with flux, although the index varies between 1.3 - 2.0 a change of 3.5$\sigma$ (see Table~\ref{tab:xspectra})

We also grouped the seven 2009 spectra into low flux (11080, 11084, 11085) and high flux (11082, 11083, 11086) subsets. We performed simultaneous fits on the members of each subset, however this yielded indistinguishable results for absorption and spectral index parameters ($N_H=0.5, \Gamma=1.5$), with poor fits in both subsets ($\chi ^2$ $>$ 2.3). 

The variation in spectral index (Table~\ref{tab:xspectra}) mirrors the hardness ratio behavior seen in Figure~\ref{fig:longlc}, confirming the picture of an event lasting perhaps weeks-months, beginning some time before November 2009, and concluding during the monitoring series in early 2010.

\section{Conclusions}
\label{sect:conclusions}
We have used  \chandra ~and \xmm ~observations spanning 2003-12 to study the light-curve and spectral changes in the black-hole X-ray binary IC 10 X-1. These observations provided information on the long-term stability of the X-ray emission, and enabled the eclipse duration to be measured. The X-ray eclipse minimum lasts for 5 hours, which combined with the previously reported mass function, leads to a larger than expected radius for the X-ray shadow cast by the stellar companion. X-1's luminosity is typically 7$\times$10$^{37}$ \lx outside of eclipse, but during 2009 exhibited an interval of slightly lower luminosity coinciding with a hardening of the spectrum. 

The eclipse features in the long-term X-ray light-curve were used to phase-connect the entire dataset leading to a refinement of the orbital period. Only a small number (4) of distinct candidate orbital periods are able to consistently and simultaneously fold the \chandra, \xmm ~and optical RV data. The most consistent X-ray period is 1.45175(1) d, which is the {\it  only} such period that places all low-flux values inside eclipse, {\it and} all high flux values outside of eclipse. Continued X-ray monitoring to capture additional eclipses will break the degeneracy between the candidate binary periods identified in this study. The ephemeris will then be known to an accuracy of $\frac{\delta P} {P} < 8 \times 10^{-6}$, enabling subsequent measurement of its period derivative; for example by followup optical RV measurements to track the accumulating phase-shift.  

Orbital period derivatives have now been measured for several BH-HMXBs including XTE J1118+480 ($\dot{P}$=-1.90$\pm$ 0.57 ms yr$^{-1}$), A0620-00 ($\dot{P}$=-0.60$\pm$0.08 ms yr$^{-1}$) \citep{Gonzalez2014}, and Cyg X-3 ($\frac{\dot{P}}{P}$=1.2-4 $\times$10$^{-6}$ s$^{-1}$;  \citealt{Lommen2005}). Simulations by \cite{Tutukov2013} predict $\frac{\dot{P}}{P}$ = 5.8$\times$10$^{-6}$ s$^{-1}$ for IC 10 X-1. Based on the mass loss expected for the WR companion, IC 10 X-1's accretion rate and hence X-ray luminosity should be $\frac{1}{2}$ Eddington, which is an order of magnitude higher than what is observed. They were able to reconcile the observed $L_X$ with the generally assumed mass of the system only by making the mass-loss rate lower. Since accreted matter transfers orbital angular momentum the rate of increase of the binary period scales with $\dot{M}$ and hence with the wind properties.

It is thought (See e.g. \citealt{Prestwich2007}) that the mass-donor star in IC 10 X-1 has a radius in accordance with the tabulated parameters of WR stars \cite{Langer1989}, which would be 1.7-2 \rsun for the most-massive case and 1.3-1.5 \rsun for the minimum mass case.  The measured RV amplitude of 370 km$s^{-1}$ \citep{SF} combined with this size estimate leads to a prediction of eclipse durations of approximately an hour in either of the mass-ratio scenarios considered (Table~\ref{tab:BHHMXBs}). There is a (factor of 5) discrepancy between that prediction and the eclipse profile in Figure~\ref{fig:eclipse} showing the X-ray shadow to be much larger  than can be accounted for by the star alone. Further refinements to the orbital model will not reduce the radius value significantly, since the corrections will mostly go in the other direction. For example if the eclipse is partial due to a lower inclination, then the chord length traversed during eclipse becomes a smaller fraction of the stellar diameter, requiring the star to be bigger. Similarly the projected velocity would be a smaller fraction of the orbital velocity, requiring the star to be larger and also more massive. Although we have ignored tidal distortion of the companion, the effect is again to underestimate the size of the star, since we only measured the projected radius in its shortest orientation. The initial radii of the progenitor stars was certainly much larger than the 1.5-2 \rsun typical of WR stars. \cite{deMink} show that the progenitors of BH-HMXB systems were larger than the present day binary separations.  As a comparison \cite{AQ} gives the radius of a 35 \msun, O supergiant  as 19 \rsun, which is close to the orbital separation we measured for X1.

Compton scattering and absorption in the wind of the secondary are expected to be significant factors in shaping the eclipse profile, given the mass loss rate is $10^{-6}-10^{-5}$ \msun $yr^{-1}$.  In the O-star binary M33 X-7, passage of the BH through the dense wind at ingress/egress extends the eclipse by $\sim$10\% due to absorption and scattering \cite{Orosz2007}. In comparison for IC 10 X-1 the WR star's wind being much denser will lead to a stronger scattering/opacity feature, as observed here. 

A very broad X-ray  modulation is also seen in LMC X-1 (which is not thought to be an eclipsing system) where the flux variation is energy independent suggesting electron-scattering as the dominant opacity mechanism  \citep{Orosz2009}. By comparison the density and velocity of the wind in IC 10 X-1 are poorly constrained, as is the degree of wind ionization, which was taken to be totally ionized in LMC X-1. The hardness ratio behavior seen for IC 10 X-1 in Figure~\ref{fig:HR_profile} and the steeper/narrower eclipse seen by \chandra ~vs \xmm ~both indicate the eclipse light-curve is strongly energy dependent.  However the general approach explains the extended duration of the eclipses in terms of wind scattering. An extended hard corona and the presence of neutral hydrogen in the eclipsing material can both produce an energy dependent eclipse edge.

An intriguing possibility is that the steep edges seen in the Chandra eclipse profile for IC 10 X-1 mark the X-ray shadow of a shock or other density enhancements in the WR-star's wind. WR winds are known to be highly asymmetric, although their termination shocks lie at $\sim$1 pc \citep{Eldredge2007}.

\begin{deluxetable}{llllll}
\tablecaption{X-ray Spectral Parameters for IC10 X-1 \label{tab:xspectra}} 
\tablehead{ \colhead{ObsID} & \colhead{$N_H$}                                            & \colhead{$\Gamma$} & \colhead{norm}                          & \colhead{$Flux$}           & \colhead{$\chi^2 / DoF$}\\
                                               &    \colhead{$10^{21} cm^{-2}$ }                       &                                    &   \colhead{ $\times10^{-5}$ }     &  & \colhead{}}
\startdata
03953 & 5.4 $\pm$ 0.3  &   1.81 $\pm$ 0.07  &   38.5 $\pm$ 2.85   &1.34   &151/155 \\
07082 &  4.3 $\pm$ 0.3  & 1.77 $\pm$ 0.07  & 30.8 $\pm$ 2.33    & 1.087  & 152/136\\
08458 &  5.5 $\pm$ 0.3  &  1.90 $\pm$ 0.06  &  44.0 $\pm$ 3.13   &1.29 & 198/155\\
11080 &  5.0 *                  &  1.41 $\pm$ 0.13 &  7.00 $\pm$ 0.82    & 0.36  & 34/22\\
11081 &  5.0 *                  &  1.48 $\pm$ 0.10  &  20.7 $\pm$ 1.91   &  1.10  & 36/34\\
11082 &  2.8 $\pm$ 0.6  &  1.31 $\pm$ 0.11  &  15.9 $\pm$ 2.18    &   & 67/66\\
11082* &  5.0 *                 &  1.58 $\pm$ 0.07  &  21.8 $\pm$ 1.36    & 1.02& 77/66 \\
11083 & 3.0 $\pm$ 0.5  &  1.48 $\pm$ 0.09  &  31.0 $\pm$ 3.25     &    & 103/87\\
11083* & 5.0 *                 &  1.73 $\pm$ 0.05  &  42.0 $\pm$ 1.94    & 1.66 & 116/87\\
11084 & 5.0 *                   &  1.90 $\pm$ 0.11  &  13.5 $\pm$ 1.21    & 0.429 &  39/28\\
11085 & 5.0 *                  &  2.03 $\pm$ 0.16  &  8.61 $\pm$ 1.01     & 0.242 & 23/17\\
11086 & 4.3 $\pm$ 0.5 &  1.73 $\pm$ 0.10  &  42.0 $\pm$ 4.76    & 1.425  & 91/83\\
\enddata
\tablecomments{Starred entries denote a frozen parameter. Both free and frozen fits are reported for 11082 and 11083. $Flux$ in units of \flux in the 0.3-8 keV band}
\end{deluxetable}

The optical radial velocity measurements made of the He II line by \cite{SF} present a complex picture when combined with the X-ray eclipse ephemeris. Three trial periods are able to place RV inferior conjunction at the mid-point of the reference X-ray eclipse. However none of these periods is able to acceptably phase-connect the X-ray dataset. Conversely the acceptable X-ray eclipse derived periods all require substantial phase shifts between the epoch of mid eclipse and inferior conjunction. This eventuality was anticipated by \cite{Tutukov2013} and points to the the need for additional spectroscopic observations to confirm the true origin of the He II emission and probe the structure of the eclipsing body.   In Section~\ref{sect:RV} we presented a novel physical mechanism for the phase shift, invoking the dust pinwheel model of \cite{Monnier2011}. A shock front must form where the WR wind collides with the wind from the BH accretion disk. Consequently a density enhanced region should trail the BH in its orbit and obscure the X-ray source when it passes in front, at 0.25 in phase before inferior conjunction.  Recent hydrodynamic modeling of the WR+BH system Cyg X-3 by \cite{Okazaki2014} shows that a dense wake probably trails the BH in its orbit, and is responsible for significant scattering and obscuration induced modulation of the X-ray source. This picture is markedly similar to the one we developed in Section~\ref{sect:RV}. 

The alternative (and more likely) interpretation is that the He II line does not directly trace the motion of the WR star, and instead originates in neutral (or partially ionized) gas shielded by a persistent structure in the stellar wind. It could be that the X-rays effectively fully ionize most of the wind as in LMC X-1 \cite{Orosz2009}, except for regions in the shadow of the star or wind-density structures (of the type discussed above). The hardness-ratio feature in the eclipses points to this being true. Emission lines require the gas to have a significant neutral content while being sufficiently energized to populate higher excited states in the atoms. Such a state of partial ionization for helium can exist above the temperature at which all hydrogen is destroyed, thus full shadowing is not required. Recombination will occur at large distances from the BH, and also in regions of enhanced wind density. The hypothesis could be tested by looking for the spectral lines emitted by other atomic and ionized species. The zone of partial ionization does not appear to be on the face of the star opposite the BH, although the orbital motion will be modified by the wind velocity, and perhaps also by stellar rotation.  

Independent of which model is correct, the RV curve and X-ray eclipse are phase shifted, and interaction between the stellar wind and the BH's wind and radiation field are required to explain the offset.   

Further X-ray (eclipse timing) and optical observations  (RV and ellipsoidal modulation) are required in order answer key remaining questions about IC 10 X1, which is of fundamental importance for the study of both stellar and intermediate mass black holes.

\section{Acknowledgements} 
We acknowledge the support of SAO grant NASA-03060 from the \chandra ~X-ray Observatory, and 
we thank UMass Lowell for supporting this research. SL thanks A. Camero for useful discussions.


\begin{appendix}

\begin{deluxetable*}{lllllllllllllll}
\tablecaption{Parameters for known BH-HMXBs \label{tab:BHHMXBs}} 
\tablehead{ \colhead{Name} & \colhead{Type}   & \colhead{$R_*$}   &  \colhead{$P_{o}$}   & \colhead{$M_*$} & \colhead{$M_{BH}$}  & \colhead{$v_*sin(i)$}  & \colhead{$v_{BH}sin(i)$}   & \colhead{$i$} & \colhead{$T_E$}         & \colhead{$\Phi_E$} & \colhead{$a_1+a_2$} & \colhead{$R_S$} & \colhead{$T_P$}   & \colhead{Refs}  \\
                                              &                            &   \colhead{\rsun}                      &       \colhead{hr}   &  \colhead{\msun} &      \colhead{\msun}    &  \colhead{km/s}         &   \colhead{km/s}      & \colhead{\degree}             & \colhead{hr}                & \colhead{phase}     & \colhead{\rsun}         & \colhead{\rsun}    & \colhead{hr}       &        } 
\startdata
IC 10 X-1 (1)  &  WR     &  2        &  34.9     &  35   &  32         &  370   &  404.7           &  90      &  5.24               &   0.15      &   22.29     &  10.5   &   1.00     &  a,b,c  \\
IC 10 X-1 (2)  &  WR    & 1.5       & 34.9      & 17    &  23         &  370   &  273.5           &  90      &   5.24              &   0.15      &    18.52    &   8.7    &  0.90     & a,b,c  \\
IC 10 X-1 (3)  &  WR    &  2         &  34.9     & 35    &  32         &  370   &  404.7           &  70     &   5.24               &   0.15      &   23.72     &   11.2  &   0.94     &  a,b,c  \\
IC 10 X-1 (4)  &  WR    &  1.5       &  34.9    & 17     &  23        &  370   &  273.5           &  70      &   5.24               &   0.15     &     19.70   &   9.3    &   0.85     & a,b,c  \\
NGC 300 X-1 &  WR    &  1.8       &  32.0    & 26     &  20        &  267   &  347.1            &   65     &    6.40             &   0.20     &   17.86      &  11.2  &   1.03     & d,e  \\
M33 X-7         &  O        &  19.6    &  82.8    & 70     &  15.65  & 108.9  &  487.1           &   74.6   &  12.19             &   0.15     &   42.17    &  19.5   &    12.25  & f     \\
LMC X-1        &  O         &  17      &  93.8    & 32     &  11        & 71.6   &  208.3            &   36.4   &  not eclipsing   &  NA     &   36.45      &      -     &    -            & g    \\
LMC X-3        &  B5V     &   4.5    &  40.92  & 3.7    &  6.95     & 240   &  127.77           &   69.6  &  not eclipsing    & NA     &  13.23        &      -     &      -         & h,i    \\
Cyg X-1         &  O         &   16.2   & 134.37 & 14.8   &  19.2    &  70    &  53.958            &   27.1  &   not eclipsing   &  NA    &   30.12      &      -     &      -         &  j   \\
NGC 4490     &  WR?   &   1.5      &  6.4      &    -     &  -           &  -      &    -                    &   -         &  not eclipsing?  &   NA   &       -           &     -      &     -          &  k   \\
Cyg X-3          & WR     &    1          &  4.8      &  10.3   & 2.4       &    109    &  469            &   43          &    not eclipsing?   &  NA     &    3.35      -       &     -     &      -          & l,m

\enddata
\tablecomments{ A ``most massive'' and ``least'' massive scenario are considered for IC 10 X-1, along with two cases for the inclination. References: 
(a)	\cite{SF}						
(b)	\cite{Prestwich2007} 						
(c)	this work						
(d)	\cite{Carpano2007}						
(e)	\cite{Crowther2010}						
(f)	\cite{Orosz2007}						
(g)	\cite{Orosz2009}						
(h)	\cite{ValBaker2007}						
(i)	\cite{Orosz2014}						
(j)	\cite{Orosz2011}						
(k)	\cite{Esposito2013}
(l)      \cite{Zdziarski2013}
(m)  \cite{Singh2002}
The following are computed from reported data: $v_{BH}sin(i)$, $a_1+a_2$, $R_S$ (Radius of X-ray shadow), $T_P$ (Predicted duration of eclipse by companion) }			
\end{deluxetable*}

We performed calculations of the type presented in Section~\ref{sect:orbitcalc} for IC10 X-1 and the known BH-HMXBs, whose properties are summarized below; the results (measured eclipse durations versus predicted) are shown in Table~\ref{tab:BHHMXBs}.\\

NGC 300 X-1 is a WR BH-HMXB with orbital period  $P_{orbit}$=1.33d (32 hrs), and an X-ray eclipse duration of about 0.2 in phase (6 hrs) as seen in figure 5 of \cite{Carpano2007}. Our inferred radius for the shadowing body (11 \msun) is very similar to that for IC 10 X-1. The observed X-ray shadow is again $\sim$5 times larger than the eclipse duration expected for the companion. Close similarity between the two systems' mass, mass ratio, companion type and X-ray spectrum have been noted (e.g. \citealt{Crowther2010}). \\

M33 X-7 is a supergiant BH-HMXB \cite{Orosz2007} with an orbital period of $P_{orbit}$=3.45d (82.8 hrs) having 12 hr long eclipses (0.147 phase duration). Unlike the WR systems in Table~\ref{tab:BHHMXBs} its X-ray eclipse duration closely matches the anticipated radius of its O supergiant companion. \\

CXOU J123030.3+413853 in NGC4490 is a candidate WR BH-HMXB \citep{Esposito2013} with a very short orbital period of just 6.4 hrs (0.267d). Its X-ray light-curve appears more sinusoidal than step-like, suggesting it is not eclipsing. The companion has not been identified owing to its great distance, so no RV curve is available, however given the short orbital period (and consequently small separation) the inclination must be very low to avoid eclipses. \\

LMC X-1 is a non-eclipsing supergiant system. Its  X-ray light curve is nonetheless strongly modulated, and is close to a cosine function with respect to orbital phase \citep{Orosz2009}.\\

LMC X-3 is a non-eclipsing system with a B3V companion. Its relatively high inclination implies that the wind does not present the wide high-opacity/scattering shadow seen in the WR (and to a lesser extent the supergiant) systems.\\

Cyg X-3  is a non-eclipsing WR BH-HMXB with a 4.8 hr (0.2 d) orbit showing strong sinusoidal-like X-ray modulation. Extensive X-ray observations by \cite{Singh2002} reveal the orbital period derivative is $\frac{\dot{P}}{P}$ = 1.05$\times$10$^{-6}$ yr$^{-1}$.

\end{appendix}

\end{document}